\documentclass[comsoc]{IEEEtran}
\usepackage{lipsum}
\usepackage{stfloats}
\usepackage{amsfonts}
\usepackage{balance}
\usepackage{fancyhdr}
\usepackage{graphicx,times,cite,amssymb,epsfig,amsthm,color}
\usepackage{multicol}
\usepackage[normalem]{ulem}
\usepackage{xcolor}
\usepackage{diagbox}
\usepackage{makecell}   
\usepackage{multirow}   

\newlength{\figwidth}
\renewcommand{\baselinestretch}{1}    
\normalsize

\usepackage{graphicx}
\ifCLASSINFOpdf
\else
\fi
\usepackage{amsmath}
\usepackage[cmintegrals]{newtxmath}
\usepackage{algorithm}
\usepackage{algorithmic}

\usepackage{array}

\usepackage{stfloats}
\usepackage{url}

\usepackage{verbatim}

\makeatletter
\newcommand*{\rom}[1]{\expandafter\@slowromancap\romannumeral #1@}
\makeatother

\begin{document}
%
\title{Constant Modulus Waveforms for IoT-Centric Integrated Sensing and Communications}

%
%
%

\author{Tian Han, Shalanika Dayarathna, Rajitha Senanayake, Peter Smith,\\
 Aryan Kaushik, Alain Mourad, Richard A. Stirling-Gallacher, Jamie Evans

        
\thanks{Tian Han (corresponding author), Shalanika Dayarathna, Rajitha Senanayake and Jamie Evans are with the Department of Electrical and Electronic Engineering, University of Melbourne, Australia; Peter Smith is with the School of Mathematics and Statistics, Victoria University of Wellington, New Zealand; Aryan Kaushik is with RakFort, Ireland, and IIITD, India; Alain Mourad is with InterDigital Europe Ltd, UK; Richard A. Stirling-Gallacher is with Huawei Technologies Duesseldorf GmbH, Germany.}
}

\markboth{Accepted to be published by IEEE Communications Standards Magazine}%
{Shell \MakeLowercase{\textit{et al.}}: Bare Demo of IEEEtran.cls for IEEE Communications Society Journals}
%



\maketitle

\begin{abstract}
Integrated sensing and communications (ISAC) is considered a key enabler to support application scenarios such as the Internet-of-Things (IoT) in which both communications and sensing play significant roles. Multi-carrier waveforms, such as orthogonal frequency division multiplexing (OFDM), have been considered as good candidates for ISAC due to their high communications data rate and good time bandwidth property for sensing. Nevertheless, their high peak-to-average-power-ratio (PAPR) values lead to either performance degradation or an increase in system complexity. This can make OFDM unsuitable for IoT applications with insufficient resources in terms of power, system complexity, hardware size or cost. This article provides IoT-centric constant modulus signalling scheme designs that leverage the advantage of unit PAPR and thus are more suitable in resource-limited scenarios. More specifically, several single-carrier frequency and/or phase signalling schemes are considered. A comprehensive discussion on their radar sensing and communications performance is conducted based on performance metrics including the radar ambiguity function, the bandwidth property, the data rate, and the communications receiver complexity. Results demonstrate that under the constraint of unit PAPR, the sensing-communications tardeoff can be achieved by selecting among the discussed signalling schemes. Recommendations for linking particular low-rate IoT scenarios with those signalling schemes are also provided based on the their performance.
\end{abstract}


%
\IEEEpeerreviewmaketitle

\section{Introduction} \label{sec:intro}



\subsection{Standardisation Status}    \label{sec:standard}

High quality communications and widespread sensing are key to next generation wireless networks. They lay the groundwork for a data-driven world in which all parts are interconnected, sensed, and equipped with intelligence. In recent years, the harmonious integration of sensing and communications (ISAC) - two functionalities that were designed and developed in isolation for decades - has garnered much research attention from academia and industry. 

Standardisation plays a crucial role in developing a consistent framework for ISAC in next generation wireless networks, within which both academia and industry can work. Recently, the industry has started laying the vision for the International Telecommunication Union Radiocommunication Sector (ITU-R) International Mobile Telecommunications-2030 (IMT-2030) towards the sixth generation of wireless communications technology (6G). ISAC has emerged as a popular theme in this vision, which led to its adoption as one of the six usage scenarios in ITU-R IMT-2030 \cite{kaushik2024isac}. This interest was the culmination of the progress made in researching orthogonal frequency division multiplexing (OFDM)-based radio radar sensing during the last decade, the successful introduction of radio frequency (RF) sensing in the Institute of Electrical and Electronics Engineers (IEEE) 802 standards at both low and high frequencies, and the enhancements made for the positioning framework in the Third Generation Partnership Project (3GPP). This positioning work is part of 5G New Radio (NR) to enable centimetre-level positioning as required by certain vertical industries.
Following ISAC adoption in ITU-R IMT-2030, in late 2023, the European Telecommunications Standards Institute (ETSI) launched an industry specification group (ISG) on ISAC with the mission to streamline ISAC pre-standardisation research and pave the way for its specification in 6G standards. In parallel, 3GPP Technical Specification Group (TSG) on Service and System Aspects Working Group 1 (SA1) launched a feasibility study on ISAC use cases in 5G-Advanced \cite{kaushik2024isac}. 
Later in 2024, 3GPP TSG Radio Access Network Working Group 1 (RAN1) started a study item focused on extending the 3GPP baseline channel model in Technical Report (TR) 38.901 to support ISAC and hence pave the way for ISAC specification in both 5G-Advanced and 6G releases. At the recent 3GPP 6G workshop in March 2025, ISAC was confirmed as one of the key themes proposed by the industry for 3GPP to specify in the forthcoming Release 21 (2028-2029) \cite{Telcotv24}. In addition to the ongoing work of the standards development organisations, extensive pre-standardisation work for important 6G ISAC use cases has also been conducted and published by various important industrial associations. These include the 5G Automotive Association \cite{5GAA}, 5G Alliance for Connected Industries and Automation \cite{5GACIA} and one6G \cite{one6G24}, who have presented various use cases in industrial and/or automated vehicle scenarios, with details of their assumptions, functional and performance requirements. 

In the development of the 6G framework, massive and sustainable Internet of Things (IoT) is considered as one of the key usage services for mobile users. ISAC is able to fulfil the sustainability aspect very well since it supports resource sharing between sensing and communication operations. For example, ISAC could allow transmitting a single joint signal for both sensing and communications tasks. This enables the reuse of spectrum, transmitter hardware and signal processing, which in turn reduces the size, complexity and power consumption of devices. Such efficient resource utilisation can improve the sustainability and reduce the cost of IoT networks, especially those with massive end devices. 




\subsection{Waveform Design}
To achieve a full integration of communications and sensing, several open challenges remain unsolved, including the important issue of joint waveform design. 
In communications, the goal is to transmit data from one place to another with a certain reliability. Thus, communication waveforms modulate the phase, frequency and amplitude (or a combination of those) of a signal in order to carry information. In sensing, more particularly in radar sensing, the goal is to detect targets and to estimate distance, velocity, and other physical state information of the target. Thus, radar waveforms are designed to have good autocorrelation properties in order to better sense the environment using the received echo. 
Integration of communications and sensing requires a joint design of new waveforms that can perform both of these functionalities.





Various approaches have been taken to tackle this joint waveform design problem \cite{Kaushik25}. Sensing (or radar) centric methods focus on embedding information into traditional radar waveforms, including linear frequency-modulated (LFM), stepped frequency or frequency hopping (FH) multiple-input multiple-output (MIMO) waveforms. Communications-centric designs integrates the radar sensing functionality into existing communications systems. Examples include designing preambles of IEEE 802.11-based frameworks and using multi-carrier waveforms such as OFDM or orthogonal time frequency space (OTFS) for sensing. Waveforms designed from the ground up do not rely on existing radar or communications waveforms. These designs are usually formulated as optimisation problems which consider both communications and sensing performance metrics 
by allocating resources among temporal, spectral and/or spatial domains. It can be observed from existing methods that the time-frequency resource allocation plays a significant role in tackling the joint waveform design problem. For example, stepped frequency radar waveforms uses a single subcarrier in each time slot, while the frequencies in different time slots are different. Thus, the entire waveform occupies a large bandwidth, leading to a high range estimation resolution. Nevertheless, its multi-carrier counterparts can achieve a much higher data rate while guaranteeing the same large bandwidth. In addition, multi-carrier systems allow for trading off between communications and radar performance by adjusting the portions of subcarriers used for transmitting data and those designed for sensing. These are some of the main reasons why OFDM has been considered an important candidate for ISAC \cite{Prelcic25}.


Nevertheless, many 6G systems may not have sufficient resources in terms of power, cost and size to support advanced multi-carrier techniques with the associated circuitry and components. For example, some existing systems such as Bluetooth low energy (BLE) and long-range radio (LoRA), as well as new applications introduced by 6G, such as internet of radars (IoR), smart wearables and wireless radar sensor networks (WRSNs) require low-power, inexpensive devices \cite{Akan20}. Therefore, single carrier modulation and low PAPR waveforms are recommended for such applications that utilise high frequency bands \cite{R1-165016}
. Given that constant envelope modulation only requires simple, low-cost circuits and operates power amplifiers at the highest efficiency point, it is ideal in low-power systems where energy efficiency and cost outweigh spectrum efficiency. As a result, in this work we focus on constant modulus waveform designs suitable for ISAC.


\subsection{Contributions}
The contributions of this work are outlined as the following.
\begin{itemize}
\item We provide a comprehensive discussion on the benefits of using low PAPR waveforms. We also introduce key performance measures for both radar and communications. 
\item We introduce multiple constant modulus frequency and/or phase signalling schemes. General discussions on the performance of these waveforms are also conducted. 
\item We provide extensive numerical examples to perform a thorough comparison among the constant modulus schemes based on the introduced performance metrics. Recommendations for linking particular IoT scenarios with those signalling schemes are also provided.
\end{itemize}

Note that though we use the phrase \emph{waveform design} throughout this paper, the major focus is a sub-topic of ISAC waveform design, i.e., the design of ISAC signalling or modulation schemes. More specifically, we focus on constant modulus ISAC signallings in the time-frequency domain. Therefore, existing works on other sub-topics, such as \cite{Li25} for signal processing, will not be discussed in this paper.





\section{Key Performance Measures} \label{sec:metrics}

In this section, we first discuss the importance of having a low peak-to-average-power-ratio (PAPR) in ISAC systems from the perspective of both radar and communications. We then present the metrics used for assessing the radar and communication performance of the constant modulus ISAC waveforms.



\subsection{Importance of Low PAPR}
In radar systems, a high average transmit power is important to achieve larger receive signal-to-noise-ratio (SNR) and thereby a larger detection range. This requires high-power amplifiers used in radar systems to be designed to operate at or near the saturation region to achieve the maximum output power efficiency. On the other hand, high-power amplifiers operating near the saturation point result in non-linear behaviour. This, combined with amplifier back-off requirements, indicates that the ratio between the peak power and the average power or PAPR needs to be low to achieve efficient performance. As a result, frequency modulated continuous wave (FMCW) radar signals and many other deterministic radar waveforms such as chirp, Barker, and Polyphase codes generally have a unit PAPR value making them highly power efficient. Any deviation from this unit PAPR results in an energy loss within the correlation mainlobe thereby reducing the peak response level and creating a detection loss in radar \cite{palo2020}.

Similarly, the detection efficiency of communication receivers is sensitive to the nonlinear devices used in the signal processing loop due to induced spectral regrowth. As such, the non-linear distortions due to high PAPR in the input signal reduce the transmitter power amplifier efficiency. This is one significant disadvantage in communication systems since high precision digital-to-analogue converters are needed to resolve this issue \cite{Jiang08}. For example, in OFDM systems, high PAPR can create problems such as out-of-band and in-band distortions. Out-of-band distortions affects the performance of users in adjacent channels due to an increased adjacent channel leakage ratio, while in-band distortions affect receiver performance due to high error vector magnitude. 
Therefore, low PAPR is an essential property in resource-limited ISAC applications. In addition, high PAPR leads to high battery consumption, which makes the 10-year battery life span requirement in IoT and IoR devices impracticable \cite{de20195g,Akan20}. 

Nonetheless, multi-carrier waveforms that suffer from high PAPR have received significant attention as ISAC waveforms due to their high data rate. As a result, different PAPR reduction techniques such as linearisation circuits have also been proposed for ISAC systems to reduce the performance degradation \cite{Jiang08}. 
However, such solutions require additional overheads of circuitry and increased complexity. This, in turn, increases the device cost or degrades the bit-error-rate (BER) performance due to OFDM signal distortion
, making them unsuitable for low-cost, low-complexity and low-power scenarios. For example, existing technologies such as BLE, LoRA and IoR do not have sufficient resources in terms of power, size and cost to accommodate multi-carrier modulation with complex PAPR reduction techniques. In contrast, single-carrier constant modulus waveforms have a unit PAPR by design, without requiring extra resources to maintain the property. Hence, they can offer advantages including better performance with simplified RF components and reduced size, weight and cost, making them ideal for applications where energy efficiency and cost outweigh spectrum efficiency, e.g., low rate IoT applications with massive battery-powered devices.





\subsection{Radar Sensing Performance}
The range and velocity (or equivalently, delay and Doppler) estimation capabilities of a radar waveform can be measured in terms of local accuracy and global accuracy. Focusing on the mainlobe of the delay-Doppler profile, i.e., the ambiguity function (AF), we can analyse the local accuracy, i.e., the delay resolution and the Doppler resolution. 
The Doppler resolution is determined by the overall time duration of the signal and, it remains constant if the time duration of the waveforms are constant. On the other hand, the radar delay resolution is directly related to the mean square bandwidth of the radar signal under a pulse compression system. The higher the mean square bandwidth, the higher the delay resolution. Therefore, we consider the mean square bandwidth to asses the radar local performance. Focusing on the rest of the delay-Doppler plane we can analyse the global accuracy of radar sensing. It describes the accuracy of target detection in the presence of clutter and this can be measured by false alarm rate and miss-detection probability. When the transmit power remains the same, as is the case for all the waveforms discussed in this paper, the miss-detection probability will be almost the same for a given threshold. On the other hand, the false alarm rate depends on the energy distribution of the received signal. 
This can be evaluated using the peak sidelobe level (PSL), defined as the ratio of the magnitude of the highest sidelobe peak in the AF to the magnitude of its mainlobe peak 
. If the PSL is lower, it mitigates the masking effect of nearby targets at the receiver matched filter output and reduces the false alarm rate. Therefore, we consider the AF PSL to assess the radar global accuracy.

Note the aforementioned sensing metrics are useful only when the target delay and/or Doppler estimation is the sensing task. While it is supplementary for Bluetooth or LoRA, technologies such as ultra-wideband IoT and IoR require high precision ranging. Some IoR scenarios even consider conventional monostatic radar architecture for localisation, since it can be realised easily when perfect synchronisation is unavailable \cite{Akan20}. 

\subsection{Communications Performance}
From the communication perspective, the performance can be measured in terms of throughput/data rate, reliability and feasibility of implementation. Note that the communication data rate depends on the channel and can vary significantly between different communication channels. Given that the scope of this work is limited to a review of different waveforms, we consider the maximum possible data rate achievable under a perfect channel instead of focusing on data rate under different communication channels. Therefore, we consider the number of bits that can be transmitted in each sub-pulse, irrespective of the channel, to assess the communication data rate in each waveform. We also note that the reliability of communication depends solely on the channels and, as such, has not been considered as a performance measure in this work. In terms of implementation feasibility, the communication transmitter of all constant modulus waveforms considered in this paper can be implemented using the existing OFDM index modulation (OFDM-IM) architecture. However, the communication receiver structure depends on the particular scheme and the considered data modulation technique as explained in Section \ref{sec:waveforms}. Therefore, we consider the computational complexity of the communication receiver to assess the feasibility of implementation for each modulation scheme. More specifically, we use the argument of the commonly used $\mathcal{O}(\cdot)$ complexity divided by the subpulse number as the per subpulse complexity measure.

\section{Constant Modulus Waveforms} \label{sec:waveforms}






\subsection{OFDM versus Single-Carrier Waveforms}
\begin{figure*}[t]
    \centering{\includegraphics[width=17.5cm]{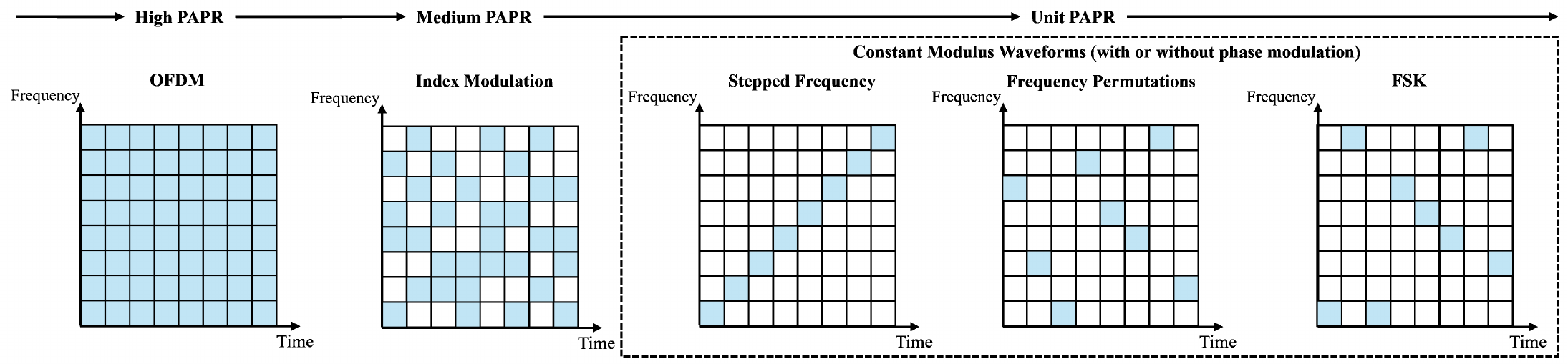}}
    \caption{ISAC waveforms with different PAPR values.}\label{fig:wfs}
\end{figure*}

While multi-carrier waveforms such as OFDM have been considered good candidates for ISAC in general, in this paper we consider constant modulus waveforms due to the importance of their unit baseband PAPR property in IoT applications. In contrast to OFDM, constant modulus waveforms use a single carrier frequency in each time slot while with no amplitude modulation, as described in Fig. \ref{fig:wfs}. While OFDM-based schemes cannot achieve a unit PAPR in general, these constant modulus waveforms can also be viewed as special cases of OFDM-IM, in which only one subcarrier is selected in each time slot.

\subsection{Waveforms with Fixed Frequency Patterns} \label{sec:fixed_freq}
Traditional stepped frequency waveforms, such as those with linear patterns of all available frequencies represented by the left-most time-frequency blocks in the dashed rectangle in Fig. \ref{fig:wfs}, provide a deterministic control of the waveform bandwidth \cite[Section 7.1.3]{Kaushik25}. 
While linear stepped frequency (LSF) waveforms suffer from high PSLs caused by large diagonal ridges in their AFs, Costas-coded stepped frequency waveforms consider well-designed frequency permutations to minimise the AF PSLs. Phase shift keying (PSK) can be independently applied to each time slot (or subpulse) of the stepped frequency waveform to embed data, which has little impact on their radar performance \cite[Section 7.1.3]{Kaushik25}. Since a fixed frequency pattern is used, the data rate and the communications receiver are just those for pure PSK. 

\subsection{Waveforms with Variable Frequency Patterns} \label{sec:var_freq}
The data modulation can also be performed using frequency modulation, for example, using all possible frequency permutations instead of a particular one \cite[Section 7.3]{Kaushik25}. An example frequency permutation is presented by the time-frequency blocks in the middle of the dashed rectangle in Fig. \ref{fig:wfs}. Such a scheme maintains the deterministic control of the waveform bandwidth, while the AF PSL varies with the frequency pattern. The data rate of the permutation-based scheme depends on the total number of waveforms, which equals the factorial of the number of subpulses. The optimal communications symbol detector uses the efficient Hungarian algorithm, whose worst-case complexity for detecting all symbols in the waveform is in the order of the number of subpulses cubed. Carefully selected subsets of frequency permutations can be considered to either improve the radar performance or decrease the communications detection error probability, though it leads to a loss of data rate \cite[Section 7.4]{Kaushik25}
. The data rate of the frequency permutation-based scheme can be boosted by incorporating subpulse-wise independent PSK with little impact on the radar performance \cite{Han2023_freqpermPSK}. However, it introduces extra complexity to the optimal communications symbol detector. 

To decrease this complexity, subpulse-wise independent frequency shift keying (FSK), shown by the right-most time-frequency blocks in the dashed rectangle in Fig. \ref{fig:wfs}, has been considered \cite{Han25global}. This allows the frequency to repeat, which means that deterministic control of waveform bandwidth is lost but rate and complexity are helped. Phase modulation can also be incorporated into the FSK scheme to either increase the data rate using PSK \cite[Section 7.2]{Kaushik25} or decrease AF PSLs using specifically designed phase sequences \cite{Han25global}. For the latter, the phase sequence for each frequency sequence is obtained by solving an optimisation problem whose objective function is the PSL. Note that with the phase sequence, the independence between subpulses no longer exists. 
In this paper, the efficient suboptimal non-coherent detector is used for the communications receiver. This detector has the same complexity as the optimal FSK detector at the cost of a slight increase in the error rate.

Please note that most conclusions on the signalling schemes mentioned in Section \ref{sec:fixed_freq} and Section \ref{sec:var_freq} are drawn from \cite[Chapter 7]{Kaushik25}, \cite{Han2023_freqpermPSK} and\cite{Han25global}. For more details, please refer to these three materials. 

To provide general guidelines on the selection of frequency and phase modulation schemes, using frequency permutations allows a higher data rate compared to fixed frequency patterns while maintains the deterministic control of the waveform bandwidth. Nevertheless, it introduces variable AF PSLs and a high communications symbol detector complexity. Using FSK reduces the communications receiver complexity and achieves a higher data rate compared to permutations, but the bandwidth varies based on the frequency pattern. In addition, combining PSK with all those mentioned frequency modulation methods further improves the data rate while have little impact on the AF PSL and the bandwidth, while the communications receiver complexity is increased. Note that the above discussion is general and qualitative since quantitative analysis and comparison are not available without specifying the waveform parameters. Hence in Section \ref{sec:comparison}, a detailed performance comparison is provided based on numerical examples.






\section{Performance Comparison}  \label{sec:comparison}



\begin{figure}[t]
\centerline{\includegraphics[width=0.5\textwidth]{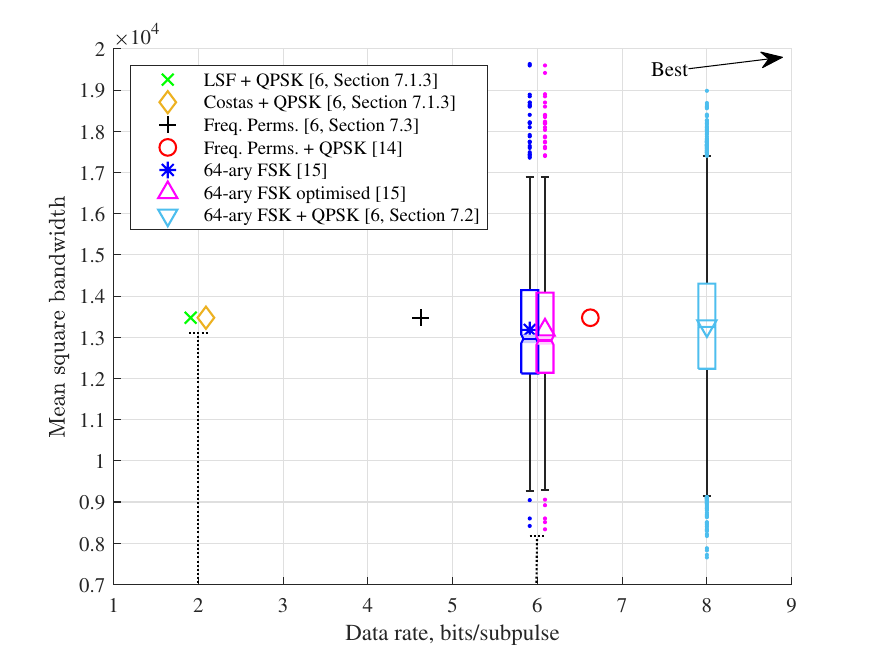}}
    \caption{Box plots of the normalised squared RMS bandwidth vs. data rate in bits/subpulse for the seven constant modulus ISAC signalling schemes. Note that we use perms. as the abbreviation of permutations in Fig. \ref{fig:msbw_vs_dr} - Fig. \ref{fig:psl_vs_rx}.} \label{fig:msbw_vs_dr}
\end{figure}

\begin{figure}[t]
\centerline{\includegraphics[width=0.5\textwidth]{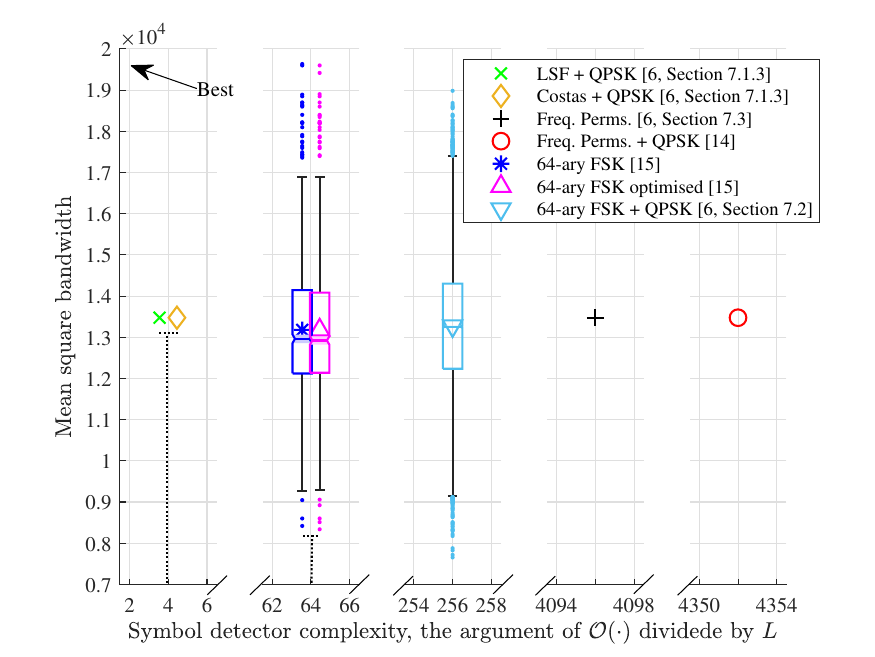}}
    \caption{Box plots of the normalised squared RMS bandwidth vs. the average per subpulse symbol detector complexity for the seven constant modulus ISAC signalling schemes.
    } \label{fig:msbw_vs_rx}
\end{figure}

\begin{figure}[h]
\centerline{\includegraphics[width=0.5\textwidth]{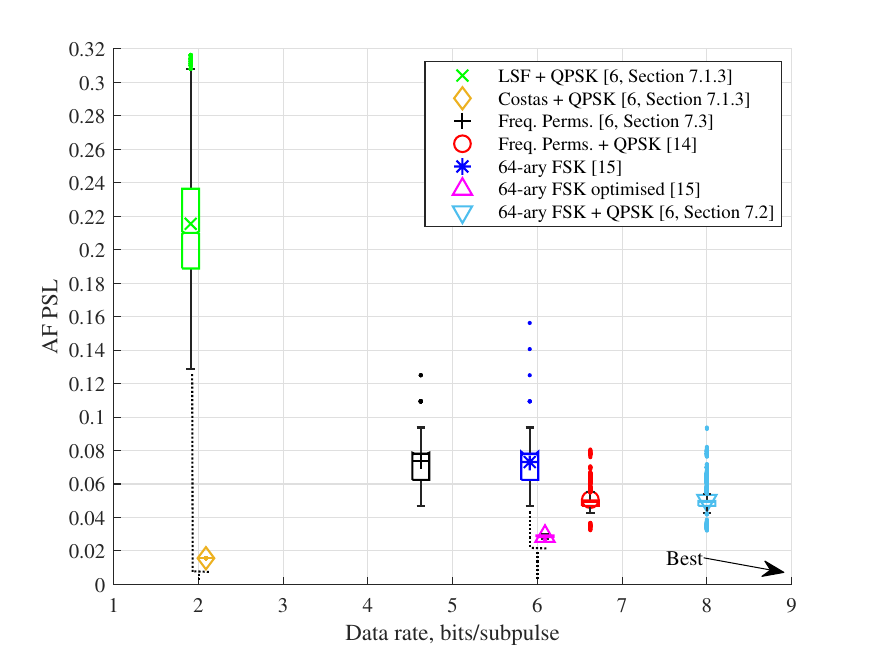}}
    \caption{Box plots of the AF PSL vs. data rate in bits/subpulse for seven constant modulus ISAC signalling schemes.} \label{fig:psl_vs_dr}
\end{figure}

\begin{figure}[h]
\centerline{\includegraphics[width=0.5\textwidth]{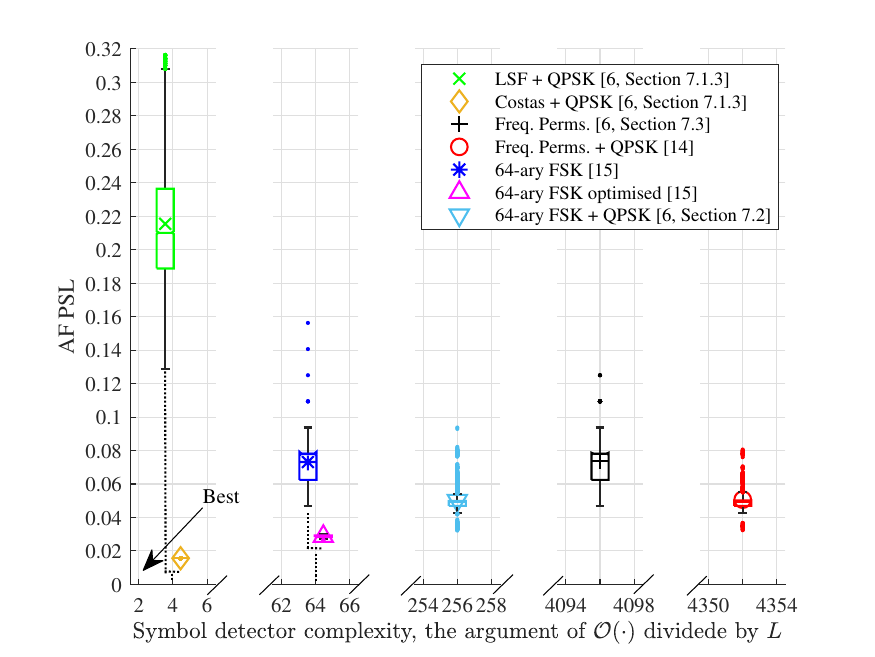}}
    \caption{Box plots of the AF PSL vs. the average per subpulse symbol detector complexity for seven constant modulus ISAC signalling schemes.
    } \label{fig:psl_vs_rx}
\end{figure}

In this section, we provide a thorough discussion of the communications and radar performance tradeoffs for all signalling schemes mentioned in Section \ref{sec:waveforms} based on the performance metrics mentioned in Section \ref{sec:metrics}. Fig. \ref{fig:msbw_vs_dr} to Fig. \ref{fig:psl_vs_rx} compare the communications and radar performance of the following schemes, all with $L = 64$ subpulses: (1) Quadrature PSK (QPSK) modulated LSF waveforms, (2) QPSK modulated Costas frequency-coded waveforms, (3) frequency permutations, (4) QPSK modulated frequency permutations, (5) $64$-ary FSK, (6) PSL-minimised $64$-ary FSK, and (7) $64$-ary FSK and QPSK modulated waveforms. 
For all schemes, we consider the bandlimited rectangular function as the subpulse shaping function with the time domain width before bandlimiting $T$ and the limiting bandwidth $B = 1/T$. The separation between adjacent frequencies for all schemes is $\Delta f = 1/T$. The selection of these waveform parameters ensures that the overall bandwidths and the time durations occupied by all schemes are the same, resulting in a relatively fair performance comparison. 

Since the theoretical analysis of some radar performance metrics is not available for all the signalling schemes considered, the radar performance metrics are numerically evaluated based on Monte Carlo simulations using $5000$ waveform realisations for each scheme. The empirical means of all radar metrics are shown by the vertical axis values of the markers labelled in the figure legends. If the value of a particular radar metric of a scheme depends on the data, its statistics are visualised using a box plot with the edge colour the same as the corresponding marker. If two schemes have the same performance for a particular communications metric, their box plots are slightly shifted horizontally to avoid overlap, with dotted lines pointing to the same value on the horizontal axes. 
In each figure, we use an arrow with the text ``Best" pointing to the corner indicating the most desired performance. 

Fig. \ref{fig:msbw_vs_dr} and Fig. \ref{fig:msbw_vs_rx} show the squared RMS bandwidth, normalised with respect to the subpulse width $T$, versus the two communications performance metrics. 
Although not shown here, the average run times of the communications detectors implemented via Matlab simulations follow the same order as the receiver complexities we presented in the numerical examples. The use of stepped frequency sequences, both the linear pattern and the Costas code, and frequency permutations results in stable delay local accuracy which is independent of the particular choice of transmitted waveform. The data rates and the receiver complexities for two PSK-modulated stepped frequency schemes depend only on the PSK modulation order, i.e., $4$, resulting in the best complexities and the worst rates among all discussed schemes; while the use of permutations leads to the second lowest data rate and a relatively high symbol detector complexity. Introducing PSK to the permutations significantly improves the data rate while not affecting the stable delay local accuracy, but it further increases the already high detector complexity. The FSK-only scheme and the phase-optimised FSK scheme have detector complexities much lower than the permutations-based ones and a relatively high data rate. Incorporating PSK with FSK (FSK+PSK) leads to the largest data rate among all schemes and a moderate receiver complexity. Nevertheless, the subpulse-wise independent frequency modulation introduces uncontrolled delay local accuracy which is shown by the outliers of these three box plots. 
We observe that the FSK-based scheme with phase modulation has similar RMS bandwidth performance to that without phase modulation. This is because the initial phase terms have a negligible impact on the RMS bandwidth of an FSK-based waveform except in the rare situation where most subpulses have the same centre frequency.

Fig. \ref{fig:psl_vs_dr} and Fig. \ref{fig:psl_vs_rx} plot the AF PSL versus the two communications performance metrics. 
Among all the signalling schemes considered, the PSK-modulated LSF scheme has the largest average AF PSL. This is caused by the huge diagonal ridge of the AF of the LSF waveform, while the phase modulation introduces a large variance to the PSLs. In contrast, the PSK-modulated Costas coded scheme has the lowest AF PSL with no variance, which is guaranteed by the perfect time-frequency correlation property of Costas codes. The FSK-only scheme also has a large average AF PSL with a large variance, due to the subpulse-wise independent frequency modulation. The restriction of using frequency permutations slightly decreases the empirical and the variance of the AF PSL. Introducing PSK modulation to both FSK and permutations decreases their average AF PSLs. Meanwhile, the majority of the AF PSL values for both PSK-modulated FSK and permutations schemes are concentrated, as can be shown by the small interquartile range and distance between whiskers. The PSL minimised FSK has an extremely low average AF PSL close to Costas with a very small variance, thanks to the solutions of the optimisation problems.

Based on the performance evaluations in Fig. \ref{fig:msbw_vs_dr} to Fig. \ref{fig:psl_vs_rx}, Table \ref{tab:rcmd} links the aforementioned signalling schemes with particular IoT scenarios. Since single carrier schemes could not provide as high data rate as their multi-carrier counterparts, in Table \ref{tab:rcmd} we only consider low to medium rate scenarios, including BLE, LoRA and IoR. LoRA and BLE require low complexity communications with low (to medium) data rates \cite{de20195g}, which can be satisfied by FSK and FSK+PSK. The uncontrolled target ranging and/or velocity estimation capability of both schemes is an insignificant issue since delay-Doppler estimation is not the main focus of LoRA and BLE. WRSNs used for target localisation prioritises high precision target sensing over communications \cite{Akan20} and thus, traditional radar waveforms-based schemes are recommended. In home scenarios with rich clutter (e.g., walls and furniture), clutter mitigation is prioritised over range resolution for sensing. The inter-device or device-smart gateway communications also require slightly higher data rate compared to WRSNs. Hence, the phase optimised FSK with close to ideal AF PSL and acceptable data rate is recommended. Special recommendations of using frequency permutation-based schemes are made for end devices that frequently perform uplink communications with the smart gateway. In this scenario, since the communications receiver is the gateway, slightly higher symbol detection complexity is acceptable.


\begin{table*}[!ht] 
\caption{Recommendations for linking IoT scenarios with signalling schemes.}
\vspace{-0.5\baselineskip}
\begin{center}
\renewcommand{\arraystretch}{1.1}
\begin{tabular}{c | c | c | c }
\hline\hline
IoT technologies & Scenarios & Performance requirements & Recommended schemes\\ 
\hline
BLE & \makecell{Short range scenarios with\\ small battery-powered devices} & \makecell{Low to medium data rate, low complexity \cite{de20195g}. \\ Target sensing is not the main focus}  &  FSK+PSK \cite[Section 7.2]{Kaushik25} \\
\hline
LoRA & \makecell{Long range scenarios \\ with massive devices} & \makecell{Low data rate, low complexity \cite{de20195g}. \\ Target sensing is not the main focus}  &  FSK \cite{Han25global} \\
\hline
\multirow{3}{*}{IoR \cite{Akan20}}  &  WRSNs for target localisation & \makecell{High precision target sensing is prioritised. \\ Very low data rate, low complexity} & Costas/LSF + PSK \cite[Section 7.1.3]{Kaushik25} \\
\cline{2-4} 
\multirow{3}{*}{}  & Clutter-limited scenarios & \makecell{Clutter mitigation is prioritised for sensing. \\ Low data rate, low complexity.} & FSK optimised \cite{Han25global} \\
\cline{2-4} 
\multirow{3}{*}{}  & \makecell{Device to smart gateway \\uplink transmission} & \makecell{Slightly higher symbol detection \\ complexity is acceptable} & \makecell{Frequency permutations \cite[Section 7.2]{Kaushik25}\\
Frequency permutations + PSK \cite{Han2023_freqpermPSK}}\\
\hline \hline
\end{tabular}
\label{tab:rcmd}  
\end{center}
\end{table*}


\section{Conclusions}
ISAC is considered a key enabler in IoT applications where both communications and sensing play important roles. Due to limits on resources, including power, system complexity, hardware size or cost in typical IoT applications, multi-carrier waveforms such as OFDM might not be suitable due to the high PAPR. To this end, this article discusses various single-carrier ISAC waveforms based on frequency and/or phase modulation schemes, including frequency permutations, FSK and PSK. These waveforms have a unit PAPR, thus making them ideal for applications where energy efficiency and cost outweigh spectrum efficiency. A comprehensive discussion on their sensing and communications capabilities is conducted based on the AF PSL, the bandwidth property, the data rate and the communications receiver complexity. Numerical examples are provided to compare their performance in detail. Results demonstrate that under the constraint of unit PAPR, phase-modulated radar frequency codes have stable sensing performance but the lowest data rates; frequency permutation-based signallings maintain the stable sensing performance and improve the data rates at a cost of the symbol detector complexity; FSK-based signallings have the highest data rate, while its AF PSLs can be controlled via phase optimisation. Based on their performance, the schemes are then recommended for particular IoT scenarios, including BLE, LoRA and IoR, where energy and cost efficiency is prioritised over spectral efficiency.

\ifCLASSOPTIONcaptionsoff
  \newpage
\fi



%




\section*{Biographies}
\begin{IEEEbiographynophoto}{Tian Han}
[Member, IEEE] (e-mail: tian.han1@unimelb.edu.au) received the B.E. degree in communication engineering from Donghua University, Shanghai, China, in 2018, the M.E. degree and the Ph.D degree in electrical and electronic engineering from the University of Melbourne, Melbourne, Australia, in 2020 and 2024, respectively. Currently, he is a research fellow with the Department of Electrical and Electronics Engineering, University of Melbourne. His main research interest is on integrated sensing and communications.
\end{IEEEbiographynophoto}

\begin{IEEEbiographynophoto}{Shalanika Dayarathna}
[S’18–M’21] (e-mail: dayarathnashalanika@yahoo.com) received the B.E. degree in electronics and telecommunications engineering from the University of Moratuwa, Sri Lanka, in 2016, and the Ph.D. degree in electrical and electronics engineering from The University of Melbourne, Australia, in 2021. From 2016 to 2018, she was with the Network Performance Assurance Team, SLT Mobitel, Sri Lanka and from 2018 to 2024, she was with the Department of Electrical and Electronics Engineering, The University of Melbourne, Australia. She is currently working as a Research Engineer at the Advances Systems and Technologies, Lockheed Martin Australia. Her research interests are in radar signal processing, resource optimization, sum-rate optimization, cooperative communications, joint radar communication and orthogonal time-frequency space modulation.
\end{IEEEbiographynophoto}

\begin{IEEEbiographynophoto}{Rajitha Senanayake}
[Member, IEEE] (e-mail: rajitha.senanayake@unimelb.edu.au) received the B.E. degree in electrical and electronics engineering from the University of Peradeniya, Sri Lanka, in 2009, the bachelor’s degree in information technology from the University of Colombo, Sri Lanka, in 2010, and the Ph.D. degree in electrical and electronics engineering from the University of Melbourne, Australia, in 2015. From 2015 to 2016, she was with the Department of Electrical and Computer Systems Engineering, Monash University, Australia. Currently, she is a senior lecturer at the Department of Electrical and Electronics Engineering, University of Melbourne. Her research interests include distributed antenna systems, fluid antenna systems and joint communications and sensing. She was a recipient of the Australian Research Council Discovery Early Career Researcher Award.
\end{IEEEbiographynophoto}

\begin{IEEEbiographynophoto}{Peter J. Smith}
[M'93, SM'01, F'15] (e-mail: peter.smith@ecs.vuw.ac.nz) received the B.Sc degree in Mathematics and the Ph.D degree in Statistics from the University of London, London, U.K., in 1983 and 1988, respectively. From 1983 to 1986 he was with the Telecommunications Laboratories at GEC Hirst Research Centre. From 1988 to 2001 he was a lecturer in statistics at Victoria University of Wellington, New Zealand. From 2001-2015 he worked in Electrical and Computer Engineering at the University of Canterbury. In 2015 he joined Victoria University of Wellington as Professor of Statistics. He was elected a Fellow of the IEEE in 2015 and in 2017 was awarded a Distinguished Visiting Fellowship by the UK based Royal Academy of Engineering at Queens University Belfast. His research interests include the statistical aspects of design, modeling and analysis for communication systems, especially antenna arrays, MIMO, mmWave systems, reconfigurable intelligent surfaces, fluid antennas and the fusion of radar sensing and communications.
\end{IEEEbiographynophoto}

\begin{IEEEbiographynophoto}{Aryan Kaushik}
[Member, IEEE] (e-mail: a.kaushik@ieee.org) is CEO at OUF Innovative, UK, since 2026, CIO at RakFort and Adjunct Professor at IIITD, since 2025. He has been Associate Professor at Manchester Met (2024-25), Assistant Professor at UoS (2021–24), Research Fellow at UCL (2020–21), and has been associated with the University of Edinburgh (2015–19), HKUST (2014–15), etc. He has held visiting appointments at Imperial College London, etc., and External PhD Examiner at KTH, UC3M, etc. He has been an invited panel member at the UK EPSRC ICT Prioritisation Panel, Chair of IEEE ComSoc ETI on ESIT and SIG on AITNTN, Vice Chair of one6G WG3, Editor of five books, etc., and has been invited as a Keynote/Tutorial speaker for 140+ academic and industry events globally, and has chaired in organizing/TPCs of 15+ flagship IEEE conferences.
\end{IEEEbiographynophoto}

\begin{IEEEbiographynophoto}{Alain Mourad}
(e-mail: alain.mourad@interdigital. com) holds a French Engineering degree in Computer Science and Telecommunications from the University of Rennes 1, France, and a PhD in Telecommunications from IMT Atlantique, France. He is a Senior Director and Head of Wireless Labs Europe at InterDigital, where he leads advanced wireless research and innovation with a strong focus on global standardisation. His work spans key international standards bodies, including ETSI, 3GPP, DVB, ATSC, IEEE 802, and the IETF, shaping the evolution of next-generation wireless technologies. Throughout his career, he has held various leadership roles at leading global technology companies, including InterDigital, Samsung Electronics, and Mitsubishi Electric, and in key international forums such as ETSI and DVB. He is widely recognized for his contributions to innovation and standards leadership, having received multiple industry and corporate awards. Dr. Mourad currently serves on the ETSI Board and as Chair of the ETSI ISAC Industry Specification Group.
\end{IEEEbiographynophoto}

\begin{IEEEbiographynophoto}{Richard A. Stirling-Gallacher} 
[Member, IEEE] (e-mail: richard.sg@huawei.com) received the M.Eng. degree in electronic engineering from the University of Southampton, U.K., in 1990, and the Ph.D. degree from The University of Edinburgh, U.K., in 1997. From 1997 to 2012, he was a Principal Researcher with Sony Deutschland GmbH, Stuttgart, Germany, where he led wireless system and algorithmic research for 3G, 4G, and millimeter-wave (mm-wave)-based communication systems. From 2012 to 2014, he was with Samsung Research America, Dallas, TX, USA, where he led research and pre-standardization for mm-wave and massive multi in multi out (MIMO) for 5G NR. He joined Huawei Technologies Inc., San Diego, CA, USA, in 2014, and subsequently in 2017, transferred to the Munich Research Center, Huawei Technologies Duesseldorf GmbH, Munich, Germany, as a Research Expert/Team Leader. He currently holds more than 140 granted U.S. patents and serves as vice chair for the ETSI ISG on ISAC. His current interests include ISAC, positioning, massive MIMO, and V2X for 6G communication systems.
\end{IEEEbiographynophoto}

\begin{IEEEbiographynophoto}{Jamie S. Evans} 
[Senior Member, IEEE] (e-mail: jse@unimelb.edu.au) received the B.S. degree in physics and the B.E. degree in computer engineering from the University of Newcastle, in 1992 and 1993, respectively, where he received the University Medal upon graduation. He received the M.S. and the Ph.D. degrees from the University of Melbourne, Australia, in 1996 and 1998, respectively, both in electrical engineering, and was awarded the Chancellor's Prize for excellence for his Ph.D. thesis. From March 1998 to June 1999, he was a Visiting Researcher in the Department of Electrical Engineering and Computer Science, University of California, Berkeley. Since returning to Australia in July 1999 he has held academic positions at the University of Sydney, the University of Melbourne and Monash University. He is currently a Professor of Electrical and Electronic Engineering and Pro Vice-Chancellor (Students and Education) at the University of Melbourne. His research interests are in communications theory, information theory, and statistical signal processing with a focus on wireless communications networks.
\end{IEEEbiographynophoto}

\end{document}